\documentclass[aps,amsmath,amssymb,superscriptaddress,prb,reprint,twocolumn,longbibliography]{revtex4-2}
\usepackage{braket,bm}
\usepackage{float}
\usepackage{mathptmx}
\DeclareMathAlphabet{\mathcal}{OMS}{cmsy}{m}{n}
\usepackage[dvipdfmx]{graphicx}
\usepackage{here}

\begin{document}
	\title{Topological Charges, Fermi Arcs, and Surface States of $K_4$ Crystal}
	\author{Shoya Yoshida}
    \affiliation{
		Department of Nanotechnology for Sustainable Energy,
		School of Science and Technology, Kwansei Gakuin University,
		Gakuen-Uegahara 1, Sanda, 669-1330, Japan
	}%

    \author{Katsuhiro Takahashi}%
    \affiliation{
		Department of Nanotechnology for Sustainable Energy,
		School of Science and Technology, Kwansei Gakuin University,
		Gakuen-Uegahara 1, Sanda, 669-1330, Japan
	}%

	\author{Katsunori Wakabayashi}%
    \affiliation{
		Department of Nanotechnology for Sustainable Energy,
		School of Science and Technology, Kwansei Gakuin University,
		Gakuen-Uegahara 1, Sanda, 669-1330, Japan
	}%
    \affiliation{
		Center for Spintronics Research Network (CSRN), 
        Osaka University, Toyonaka 560-8531, Japan 
	}%
    \affiliation{
		National Institute for Materials Science (NIMS), 
        Namiki 1-1, Tsukuba 305-0044, Japan
	}%
	\date{\today}
\begin{abstract}
We investigate the topological electronic properties of the $K_4$ crystal 
by constructing a tight-binding model. 
The bulk band structure hosts Weyl nodes with higher and conventional chiralities 
($\chi = \pm 2$ and $\chi = \pm 1$) located at high-symmetry points in the Brillouin zone. 
Through analytical evaluation of the Berry curvature, we identify the positions 
and chiralities of these Weyl nodes. 
Furthermore, slab calculations for the (001) surface reveal Fermi arcs 
that connect Weyl nodes of opposite chirality, 
including those linking $\chi = \pm 2$ nodes with pairs of $\chi = \mp 1$ nodes. 
These results demonstrate that the $K_4$ crystal is a spinless Weyl semimetal 
featuring topologically protected surface states originating from 
multiple types of Weyl nodes.
\end{abstract}

\maketitle

\section{Introduction}
In recent years, the study of topological materials-including
topological insulators~\cite{Kane2005,Bernevig2006,Fu2007,Konig2007,
Qi2011TIandSC,Hsieh2008,Chen2009,Chang2013,Ando2013,
Hasan2010colluquim,qi2010,zhang2010crossover-65c},
topological crystalline
insulators~\cite{fu2011topological-58b,hsieh2012topological-792},
topological
magnets~\cite{tokura2019magnetic-87b,mong2010antiferromagnetic-9f7}, and
Dirac and Weyl
semimetals~\cite{Wan2011TSandFermiArcs,Xu2015DiscoveryWeylFermion}-has
revealed novel quantum phases characterized by topological invariants
and robust surface states~\cite{Qi2011TIandSC, Wang2012DSMA3Bi}. 
Among these systems, Weyl semimetals are distinguished by pairs of
topologically protected band crossings, known as Weyl nodes, which act
as monopoles of Berry curvature in momentum
space~\cite{Wan2011TSandFermiArcs,Armitage2018WeylandDirac}. 
A hallmark feature of Weyl semimetals is the emergence of open surface states,
called Fermi arcs, which connect projections of Weyl nodes with opposite
chirality in the slab Brillouin zone~\cite{Xu2015DiscoveryWeylFermion}. 
These arcs directly reflect the nontrivial bulk topology and serve as
unambiguous experimental signatures of Weyl physics~\cite{Xu2015DiscoveryWeylFermion,
Deng2016Experimental}. 
Understanding how Fermi arcs emerge and how their connectivity is dictated
by the underlying crystal symmetry has become a central theme in the
exploration of topological materials. 

The theoretical design of novel crystal structures, 
inspired by mathematical principles not realized in nature,
has attracted considerable attention in recent years. 
One prominent example is the $K_4$ lattice, 
a mathematical construct defined as the maximal abelian covering graph 
of the complete graph $K_4$~\cite{sunada2008CrystalThatNature,sunada2013topological-8e2}. 
Owing to its unique geometry, 
this structure has been referred to by various alternative names,
including the diamond twin~\cite{sunada2019diamond}, the srs
lattice~\cite{sunada2013topological-8e2,sunada2012lecture-da8,
okeeffe2008reticular-5ef}, and the Laves
graph~\cite{sunada2013topological-8e2,hyde2008short-293}. 
It is also closely related to the Gyroid
surface~\cite{sunada2013topological-8e2, hyde2008short-293}, a chiral
network found in nature-for instance, in butterfly wing
scales~\cite{michielsen2008gyroid-6fd}. 
Although the $K_4$ lattice has
not yet been directly observed in natural systems, the existence of
structurally analogous biological networks suggests that such
mathematical constructs can serve as blueprints for the design of novel
materials in both synthetic and biological contexts. 

In condensed matter physics, the $K_4$ crystal has attracted considerable
interest because its unique topology and electronic structure are
expected to give rise to unprecedented physical
properties~\cite{koretsune2009new,Tsuchiizu2016ThreeDimensional,hagita2016dft-ef9}. 
Materials adopting this crystal are predicted to exhibit electronic behaviors
distinct from those of familiar carbon allotropes such as
graphene~\cite{RevModPhys.81.109,novoselov2005two-dimensional,zhang2005}
and diamond~\cite{WORT200822,PhysRevB.71.205214}. 
This highlights the potential of the $K_4$ crystal as a guiding principle 
for the design of functional materials in both solid crystals and self-assembled
architectures. 
Theoretical investigations of the $K_4$ crystal have
extended beyond carbon allotropes to a wide range of other material
systems. For example, boron and phosphorus $K_4$ crystals can form
stable sp$^2$ chiral structures~\cite{dai2010boron-234,
liu2016phosphorus-b0c}. 
Moreover, multielement $K_4$ crystals have
been proposed through doping or intercalation, including hydrogenated
$K_4$ carbon~\cite{lian2013hydrogenated-87a}, carbon-doped $K_4$
nitrogen~\cite{wen2011carbondoped-abc}, and $K_4$-like
NaC$_2$~\cite{wen2012mechanical-93d}. 
The $K_4$ crystal structure has been experimentally realized 
via electrochemical crystallization of radical anion salts 
derived from a triangular molecule~\cite{doi:10.1021/jacs.5b04180}.

One of the most striking features of the $K_4$ crystal is the emergence
of a so-called \textit{triple Dirac cone} in its electronic band
structure. 
At high-symmetry points such as $\Gamma$ and H, two linearly
dispersing bands intersect with an almost flat band, forming a
pseudospin-1 Dirac point~\cite{Tsuchiizu2016ThreeDimensional}. 
Such band crossings are rare in crystalline systems. 
\begin{figure*}[t]
    \includegraphics[width=0.9\linewidth]{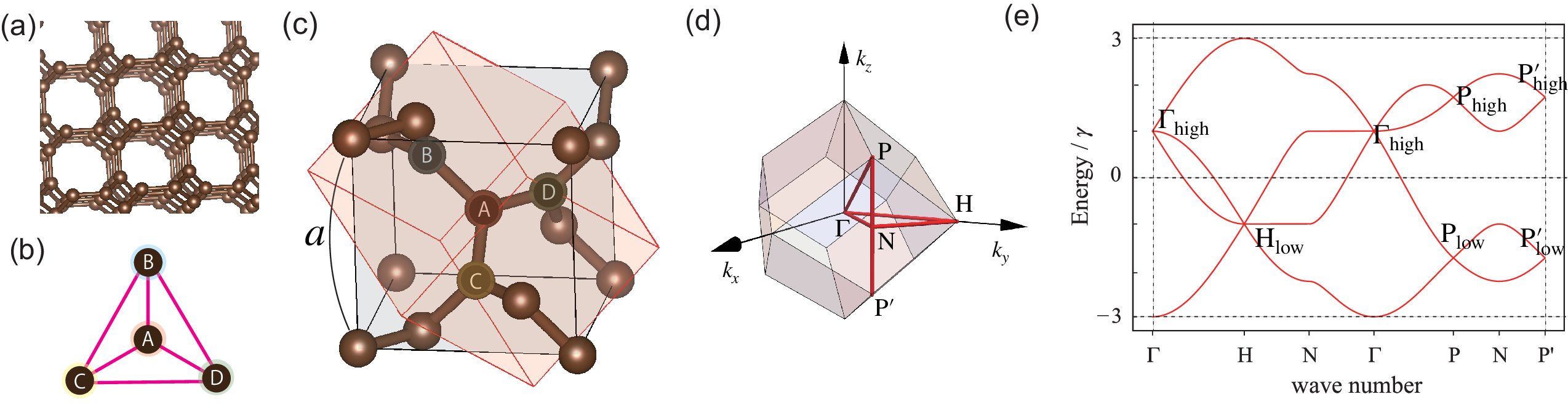}
    \caption{
        (a) Crystal structure of the $K_4$ crystal. 
        Viewed along the (001) direction, it exhibits a pattern of tiled squares and octagons
        and contains multiple helical motifs.
        (b) Graph representation of the $K_4$ structure,
        showing the connectivity of the four sublattices. 
        The corresponding graph is a complete graph with four vertices, 
        each connected to all others by three edges.
        (c) Unit cell of the $K_4$ crystal.
        The gray cube represents the conventional cubic unit cell, 
        while the red rhombohedron denotes the primitive unit cell 
        of the body-centered cubic (bcc) lattice.
        The primitive cell contains four sublattices, 
        labeled A-D, each connected by three nearest-neighbor bonds.
        At each site, three bonds extend isotropically in a single plane, 
        similar to graphene. However, the bond plane at one site 
        is rotated by an angle of $\theta \simeq 70.5^\circ$ ($\cos \theta = 1/3$) 
        relative to that of a neighboring site.
        (d) First Brillouin zone (BZ) of the $K_4$ crystal, 
        based on the bcc lattice.
        (e) Energy band structure of the $K_4$ crystal.
        At the $\Gamma$ point, two linearly dispersing (conical) bands 
        and one flat band intersect at energy $\gamma$, 
        referred to as the $\Gamma_{\mathrm{high}}$ point.
        At the H point, a similar crossing occurs at energy $-\gamma$, 
        labeled as the $\mathrm{H_{low}}$ point.
        The four-band structure around $\mathrm{H_{low}}$ 
        is the vertical inversion of that around $\Gamma_{\mathrm{high}}$.
        At the P points, two conical bands form 
        vertically aligned degenerate points at energies
        $-\sqrt{3}\gamma$ and $+\sqrt{3}\gamma$,
        denoted as $\mathrm{P_{low}}$ and $\mathrm{P_{high}}$,
 respectively.
    }\label{fig:k4_structure_band}
\end{figure*}

Beyond conventional Dirac and Weyl fermions, the discovery of multi-fold
band degeneracies has opened new directions in the study of topological
phases. Crystalline symmetries can stabilize higher-fold fermions such
as triple-point fermions, spin-1 Weyl fermions, and other exotic
quasiparticles, which in turn host unconventional topological charges
and novel transport or surface phenomena
~\cite{bradlyn2016beyond-79c,zhu2016triple-68f,
flicker2018chiral-3ea,wang2025exhaustive-e4d,schroter2020weyl-aa2,xu2020optical-495,
hu2018TopologicalTriply,fulga2017triple-2b7,yang2019topological-919}. 
These studies highlight that the triple Dirac cone in the $K_4$ crystal belongs 
to a broader framework of topological semimetals stabilized by symmetry-protected
degeneracies. 

In this work, we construct a tight-binding model of the $K_4$ crystal
using one $s$ orbital per site with nearest-neighbor hopping and analyze
the resulting band structure. 
By evaluating the Berry curvature
~\cite{vanderbilt2018berry-985, xiao2010berry-83a, zak1989berrys-6d8,
kingsmith1992theory-0ea, resta1994macroscopic-268, marzari2012maximally-8d1,
zhou2015surface-a78, liu2017novel-f2e, delplace2011zak-fd0,PhysRevB.109.035431},
we identify Weyl nodes at high-symmetry points including $\Gamma$, H, and
P, and classify them according to their chirality. 
We further
investigate the surface states in a (001) slab geometry, demonstrating
the presence of topological Fermi arcs that connect Weyl nodes of
different chiralities. 
In particular, we show that higher-chirality
($\chi = \pm 2$) Weyl nodes are linked to pairs of conventional ($\chi =
\pm 1$) nodes via topologically protected surface states. 

Thus, this study fills a crucial knowledge gap: demonstrating that the
$K_4$ crystal is not only a mathematically intriguing lattice construct or a
metallic carbon allotrope, but also a bona fide Weyl semimetal with
topologically nontrivial surface states. 
This establishes the $K_4$
crystal as a novel archetype of three-dimensional $sp^2$-hybridized
carbon with intrinsic topological properties.

\section{Crystal and Band Structure}
Figure~\ref{fig:k4_structure_band}(a) shows the crystal structure of the $K_4$ crystal.  
It can be regarded as a three-dimensional realization 
of the complete graph $K_4$, in which all four vertices are mutually connected.  
This unique connectivity is reflected in the atomic arrangement shown 
in Fig.~\ref{fig:k4_structure_band}(b), where each site forms bonds with the other three.  
As a result, three bonds emanate from each site in a coplanar configuration; 
however, the planes formed by these bonds are twisted relative to 
those of neighboring sites by an angle $\theta \simeq 70.5^\circ$ (with $\cos\theta = 1/3$).  
When all such planes are aligned parallel, the resulting structure 
reduces to the two-dimensional honeycomb lattice.  
Thus, the $K_4$ crystal can be viewed as a three-dimensional extension of the honeycomb structure.

The crystal structure of $K_4$ is derived from this connectivity of
$K_4$ graph.
It adopts a body-centered cubic (bcc) lattice
with space group $I4_132$, which lacks inversion symmetry
and thus renders the system noncentrosymmetric.
The conventional unit cell with lattice constant $a$
is shown in Fig.~\ref{fig:k4_structure_band}(c).
The rhombohedral primitive cell denoted by the red rhombohedron 
in Fig.~\ref{fig:k4_structure_band}(c) is spanned by the translation vectors
\begin{align*}
    \bm{a}_1 = \frac{a}{2}\left(-1,1,1\right), 
    \bm{a}_2 = \frac{a}{2}\left(1,-1,1\right), 
    \bm{a}_3 = \frac{a}{2}\left(1,1,-1\right),
\end{align*}
and contains four atomic sites labeled $A$, $B$, $C$, and $D$.
The fractional coordinates of the four sublattice sites are given by
\begin{align*}
    \bm{R}_A &= a\left(0 ,\; 0 ,\; 0\right),
    &\bm{R}_B = a\left(\frac{1}{2} ,\; \frac{1}{4} ,\; -\frac{1}{4}\right),\\
    \bm{R}_C &= a\left(-\frac{1}{4} ,\; \frac{1}{2} ,\; \frac{1}{4}\right),
    &\bm{R}_D = a\left(\frac{1}{4} ,\; -\frac{1}{4} ,\; \frac{1}{2}\right).
\end{align*}
Each site is connected to its three nearest neighbors by
the displacement vectors $\bm{\tau}_{ij}$ ($i,j=A,B,C,D$),
among which six are independent:
\begin{align*}
  \bm{\tau}_{AB} &= \frac{a}{4}\left(0 ,1, -1\right),
  \bm{\tau}_{AC} = \frac{a}{4}\left(-1,0 ,1\right),
  \bm{\tau}_{AD} = \frac{a}{4}\left(1, -1, 0\right),
  \\
  \bm{\tau}_{BC} &= \frac{a}{4}\left(1, 1, 0\right),
  \bm{\tau}_{BD} = \frac{a}{4}\left(-1, 0 ,-1\right),
  \bm{\tau}_{CD} = \frac{a}{4}\left(0, 1, 1\right).
\end{align*}
The corresponding reciprocal vectors are given as
\begin{align*}
    \bm{b}_1 = \frac{2\pi}{a}\left(0 , 1 , 1\right),
    \bm{b}_2 = \frac{2\pi}{a}\left(1 , 0 , 1\right),
    \bm{b}_3 = \frac{2\pi}{a}\left(1 , 1 , 0\right).
\end{align*}
Thus, the first Brillouin zone (BZ) of the bcc lattice 
has a rhombohedral shape, as shown in Fig.~\ref{fig:k4_structure_band}(d), 
with the following representative high-symmetry points:
\begin{align*}
    &\mathrm{\Gamma} : \left(0 ,\; 0 ,\; 0\right),
    &&\mathrm{H} : \left(0 ,\; \frac{2\pi}{a} ,\; 0\right),\\
    &\mathrm{N} : \left(\frac{\pi}{a} ,\; \frac{\pi}{a} ,\; 0\right),
    &&\mathrm{P} : \left(\frac{\pi}{a} ,\; \frac{\pi}{a} ,\; \frac{\pi}{a}\right),
    \mathrm{P'} : \left(\frac{\pi}{a} ,\; \frac{\pi}{a} ,\; -\frac{\pi}{a}\right).
\end{align*}

The symmetry properties of the high-symmetry points (k-point little groups) 
in the Brillouin zone are summarized as follows.
The $\Gamma$ and $\mathrm{H}$ points belong to the $O$ point group ($432$).
The $\Gamma$ point is unique under this symmetry,
while the $\mathrm{H}$ point has six symmetry-equivalent points located at
$(\pm \frac{2\pi}{a}, 0, 0)$, $(0, \pm \frac{2\pi}{a}, 0)$, and $(0, 0, \pm \frac{2\pi}{a})$.
The $\mathrm{P}$ and $\mathrm{P'}$ points belong to the $T$ point group ($23$).
Each of these has four symmetry-equivalent points related by the cubic symmetry operations:
\begin{widetext}
\begin{align*}
\mathrm{P}: &\;
    \left(+\tfrac{\pi}{a}, +\tfrac{\pi}{a}, +\tfrac{\pi}{a}\right),\;
    \left(+\tfrac{\pi}{a}, -\tfrac{\pi}{a}, -\tfrac{\pi}{a}\right),\;
    \left(-\tfrac{\pi}{a}, +\tfrac{\pi}{a}, -\tfrac{\pi}{a}\right),\;
    \left(-\tfrac{\pi}{a}, -\tfrac{\pi}{a}, +\tfrac{\pi}{a}\right), \\
\mathrm{P'}: &\;
    \left(-\tfrac{\pi}{a}, -\tfrac{\pi}{a}, -\tfrac{\pi}{a}\right),\;
    \left(-\tfrac{\pi}{a}, +\tfrac{\pi}{a}, +\tfrac{\pi}{a}\right),\;
    \left(+\tfrac{\pi}{a}, -\tfrac{\pi}{a}, +\tfrac{\pi}{a}\right),\;
    \left(+\tfrac{\pi}{a}, +\tfrac{\pi}{a}, -\tfrac{\pi}{a}\right).
\end{align*}
\end{widetext}
The $\mathrm{N}$ point is invariant only under the identity operation,
and no band degeneracy occurs there, so we omit a detailed discussion.

We construct a tight-binding model for $K_4$ crystal 
by assuming a single $s$-orbital per site and only nearest-neighbor hopping. 
The onsite energy is set to zero at all sites.
Thus, the eigenvalue equation of $K_4$ crystal reads 
\begin{align*}
 \hat{H}(\bm{k}) | u_j(\bm{k})\rangle =
 E_j(\bm{k}) | u_j(\bm{k})\rangle,
\end{align*}
where $\bm{k}=(k_x, k_y, k_z)$ is the wavenumber vector and $j(=1,2,3,4)$
is the band index. 
The eigenvector is defined as 
$|u_j(\bm{k})\rangle = (\psi_{jA}(\bm{k}),
\psi_{jB}(\bm{k}), \psi_{jC}(\bm{k}),\psi_{jD}(\bm{k}))^T$, where
$(\cdots)$ indicates the transpose of the vector.
$\psi_{j\alpha}(\bm{k})$ ($\alpha=A,B,C,D$) is the amplitude at site
$\alpha$ for $j$-th energy band at $\bm{k}$. 
The Hamiltonian in momentum space takes the form of 
a $4 \times 4$ Hermitian matrix $H(\bm{k})$, where each matrix element represents 
hopping between two sites connected by a nearest-neighbor bond. 
The off-diagonal matrix element, $h_{ij}(\bm{k})$, which means  
the electron hopping from site $j$ to site $i$, is given by
\begin{equation*}
h_{ij}(\bm{k}) = -\gamma\, e^{-i\bm{k} \cdot \bm{\tau}_{ij}},
\end{equation*}
where $i,j=1,2,3,4$. $\gamma$ is the nearest-neighbor hopping amplitude.
$\bm{\tau}_{ij}$ is the displacement vector from site $j$ to site $i$.
Since we have assumed the onsite potential is set to zero at all sites,
the diagonal matrix elements are zero, i.e., 
$h_{ii}(\bm{k}) = 0$.
Thus the resulting Hamiltonian matrix is explicitly written as
\begin{widetext}
\begin{align}
    \hat{H}(\bm{k}) 
  &= -\gamma
  \begin{pmatrix}
    0&e^{-i(k_y-k_z)\frac{a}{4}}&e^{-i(k_z-k_x)\frac{a}{4}}&e^{-i(k_x-k_y)\frac{a}{4}} \\
    e^{i(k_y-k_z)\frac{a}{4}}&0&e^{-i(k_x+k_y)\frac{a}{4}}&e^{-i(-k_z-k_x)\frac{a}{4}} \\
    e^{i(k_z-k_x)\frac{a}{4}}&e^{i(k_x+k_y)\frac{a}{4}}&0&e^{-i(k_y+k_z)\frac{a}{4}} \\
    e^{i(k_x-k_y)\frac{a}{4}}&e^{i(-k_z-k_x)\frac{a}{4}}&e^{i(k_y+k_z)\frac{a}{4}}&0 \\
  \end{pmatrix}.
  \label{eq:k4hamil}
\end{align}
\end{widetext}

Figure~\ref{fig:k4_structure_band}(e) shows the band structure 
of the $K_4$ crystal obtained from this tight-binding model.
At the band degeneracy points $\Gamma_{\mathrm{high}}$ and $\mathrm{H_{low}}$,
we find a threefold band crossing composed of two linearly dispersing bands forming a Dirac cone,
together with a nearly flat band intersecting at the same point.
This type of threefold degeneracy is termed a \textit{triple Dirac cone}.
In contrast, at the $\mathrm{P_{low}}$ and $\mathrm{P_{high}}$ points, 
a twofold degeneracy forms a conventional Dirac cone with linear
dispersion.

From the viewpoint of group theory, the 3+1 band degeneracy structure 
at the $\Gamma$ and H points is classified as the direct sum of 
irreducible representations $T_2 \oplus A_1$ of the $O$ point group.
Likewise, the 2+2 degeneracy at the $\mathrm{P}$ and $\mathrm{P'}$ points 
corresponds to the $E \oplus E$ representation of the $T$ point group.

Furthermore, the band degeneracies can be understood from the viewpoint of
Wyckoff positions~\cite{ITA_VolA_2016, wyckoff_crystal_structures} 
and elementary band representations (EBRs)~\cite{bradlyn2017topological-26b}.  
The four sublattices in the $K_4$ unit cell occupy the Wyckoff position $8a$
of the space group $I4_132$ (No.~214), whose site symmetry is $32$ 
(the $D_3$ point group).  
Table~\ref{tab:ebr} summarizes the band representations induced from the
$8a$ Wyckoff position, taken from the Bilbao Crystallographic Server
~\cite{aroyo2006bilbao1}.
\begin{widetext}
    \begin{table}[h]
        \caption{Band representations for the Wyckoff position $8a$
     $(1/8,1/8,1/8)$ of space group $I4_132$ (No.~214)}
        \centering
        \label{tab:ebr}
            \begin{tabular}{c|ccc}
                \hline\hline
                Band-Rep. 			& $A_1 \uparrow G(4)$ & $A_2 \uparrow G(4)$ & 
                    $E \uparrow G(8)$ \\
                \hline
                $\Gamma:(0,0,0)$ 	& $\Gamma_1(1) \oplus \Gamma_5(3)$ & $\Gamma_2(1) \oplus \Gamma_4(3)$ & 
                    $\Gamma_3(2) \oplus \Gamma_4(3) \oplus \Gamma_5(3)$ \\
                H:$(1,1,1)$ 		& $H_2(1) \oplus H_4(3)$ & $H_1(1) \oplus H_5(3)$ & 
                    $H_3(2) \oplus H_4(3) \oplus H_5(3)$ \\
                P:$(1/2,1/2,1/2)$ 	& $P_1(2) \oplus P_3(2)$ & $P_1(2) \oplus P_3(2)$ & 
                    $P_1(2) \oplus 2P_2(2) \oplus P_3(2)$ \\
                N:$(1/2,1/2,0)$ 	& $2N_1(1) \oplus N_3(1) \oplus N_4(1)$ & 
                    $2N_2(1) \oplus N_3(1) \oplus N_4(1)$ & 
                    $2N_1(1) \oplus 2N_2(1) \oplus 2N_3(1) \oplus 2N_4(1)$ \\
                \hline\hline
            \end{tabular}
    \end{table}
\end{widetext}

For the spinless $s$-orbital tight-binding model, the bands transform according to
the elementary band representation $A_1 \uparrow G(4)$.  
Consequently, at the $\Gamma$ and H points, this representation decomposes into
$\Gamma_1 \oplus \Gamma_5$ and $H_2 \oplus H_4$, respectively, producing the
characteristic $3+1$ band degeneracy.  
At the P point, it yields $P_1 \oplus P_3$, leading to the observed $2+2$
degeneracy. Since the calculated bands coincide with a sum of elementary
band representations (EBRs), the system is classified as being in an
atomic limit. 

When the $p_x$, $p_y$, and $p_z$ orbitals are included, the atomic degrees of freedom
generate the representation
\begin{align*}
    A_1 \uparrow G(4) \oplus A_2 \uparrow G(4) \oplus E \uparrow G(8).
\end{align*}
Here, $A_2 \uparrow G(4)$ corresponds to the $p$ orbital perpendicular to the
coplanar plane spanned by the three nearest-neighbor bonds, and
$E \uparrow G(8)$ corresponds to the remaining two in-plane $p$ orbitals.
The hybridization of $s$ and $p$ orbitals form the $\pi$ bands, in which 
triple degeneracies at $\Gamma$ and H and the double degeneracies at P
persist in the same manner as in the $s$-orbital model.

Thus, these degeneracies persist even in the multiorbital ($s+p$) system, 
as confirmed by symmetry analysis.
In addition, the same degeneracies have been reported in first-principles calculations
~\cite{koretsune2009new,Tsuchiizu2016ThreeDimensional}.
This indicates that the minimal $s$-orbital model already captures 
the essential topological features of the $K_4$ crystal.

\section{Topological Charge}
In this section, we demonstrate that the band degeneracies 
identified in $K_4$ crystal correspond to Weyl points, 
each carrying a quantized topological charge (chirality).

\subsection{Weyl points}
A Weyl point is a singularity in momentum space at which two or more energy bands 
cross linearly in all directions, giving rise to gapless excitations 
governed by the Weyl equation
~\cite{weyl1929elektron-50c, Armitage2018WeylandDirac, Wan2011TSandFermiArcs}.
Near such a point $\bm{k}_0$, the effective Hamiltonian 
can be expressed as
\begin{align}
H(\bm{q}) = \left(E_0 + \bm{v}_0 \cdot \bm{q} \right) I
+ \sum_{i = x, y, z} \bm{v}_i \cdot \bm{q} \sigma_i,
\label{eq:Weyl_eq}
\end{align}
where $\bm{q} = \bm{k} - \bm{k}_0$ is the wavevector measured from the Weyl point,
$E_0$ and $\bm{v}_0$ denote the energy and group velocity at $\bm{k}_0$,
$I$ is the identity matrix, $\sigma_i$ are the Pauli matrices,
and $\bm{v}_i$ represent the effective velocities along the $i$-axis.

Here, we explicitly construct the effective Hamiltonians around 
the degeneracy points $\Gamma_{\mathrm{high}}$, $\mathrm{H_{low}}$, 
$\mathrm{P_{low}}$ and $\mathrm{P_{high}}$, in the $K_4$ crystal
and calculate eigenvalues and eigenfunctions.
The effective Hamiltonians are obtained by expanding the tight-binding Hamiltonian [Eq.~\eqref{eq:k4hamil}] 
to first order in $\bm{k}$ and applying appropriate unitary transformations
that diagonalize the zeroth-order Hamiltonians at each Weyl point.
For related methods and eigenvalue analyses, see Ref.~\cite{Tsuchiizu2016ThreeDimensional}.

We first focus on the triple band degeneracy located at the $\Gamma_{\mathrm{high}}$ point, 
which consists of two dispersive (conical) bands and one flat band. 
At the $\Gamma$ point ($\bm{k} = \left(0 ,\; 0 ,\; 0\right)$), 
the zeroth-order Hamiltonian is given by
\begin{align*}
    H_\Gamma^{(0)}
    = -\gamma
    \begin{pmatrix}
      0&1&1&1 \\
      1&0&1&1 \\
      1&1&0&1 \\
      1&1&1&0 \\
    \end{pmatrix},
\end{align*}
where the Bloch phase factors drop out, and the four tight-binding basis states 
$|A\rangle, |B\rangle, |C\rangle$, and $|D\rangle$
are connected by equivalent hopping $-\gamma$. 
This connectivity corresponds to the $K_4$ graph (Fig.~\ref{fig:k4_structure_band}(b)) itself 
and can be viewed as a regular tetrahedron.
The representation of these four basis functions 
follows the point group $\mathrm{T_d}$,
which decomposes into the irreducible representations 
$A_1$ and $T_2$, giving rise to one nondegenerate state 
($A_1$, $E=-3\gamma$) and one triply degenerate state 
($T_2$, $E=+\gamma$) at the $\Gamma$ point.

Expanding the Hamiltonian to first order in $\bm{k}$, we obtain
\begin{widetext}
\begin{align*}
    H_\Gamma^{(1)}(\bm{k}) 
    = -\gamma
    \begin{pmatrix}
      0&1&1&1 \\
      1&0&1&1 \\
      1&1&0&1 \\
      1&1&1&0 \\
    \end{pmatrix}
    -\frac{\gamma a}{4}
    \begin{pmatrix}
      0&-i(k_y-k_z)&-i(k_z-k_x)&-i(k_x-k_y) \\
      i(k_y-k_z)&0&-i(k_x+k_y)&-i(-k_z-k_x) \\
      i(k_z-k_x)&i(k_x+k_y)&0&-i(k_y+k_z) \\
      i(k_x-k_y)&i(-k_z-k_x)&i(k_y+k_z)&0 \\
    \end{pmatrix}.
\end{align*}
This Hamiltonian can be block-diagonalized using the unitary transformation $U_\Gamma$:
\begin{align*}
    U_\Gamma^\dagger H_\Gamma^{(1)}(\bm{k}) U_\Gamma =
    &-\gamma
    \begin{pmatrix}
      3&0&0&0\\
      0&-1&0&0\\
      0&0&-1&0\\
      0&0&0&-1\\
    \end{pmatrix}
    -\frac{\gamma a}{2}
    \begin{pmatrix}
      0&0&0&0\\
      0&k_z&\frac{1}{\sqrt{2}}k_-&0\\
      0&\frac{1}{\sqrt{2}}k_+&0&\frac{1}{\sqrt{2}}k_-\\
      0&0&\frac{1}{\sqrt{2}}k_+&-k_z\\
    \end{pmatrix},
\end{align*}
\end{widetext}
where $k = \sqrt{k_x^2+k_y^2+k_z^2}$ and $k_{\pm} = k_x\pm ik_y$. 

\clearpage
The unitary matrix $U_\Gamma$ is constructed from the zeroth-order eigenfunctions 
at the $\Gamma$ point as
\begin{align*}
    U_\Gamma &= \frac{1}{2}
    \begin{pmatrix}
      1&e^{-i\frac{\pi}{4}}&e^{+i\frac{\pi}{2}}&e^{+i\frac{\pi}{4}}\\
      1&e^{-i\frac{3\pi}{4}}&e^{-i\frac{\pi}{2}}&e^{+i\frac{3\pi}{4}}\\
      1&e^{+i\frac{\pi}{4}}&e^{-i\frac{\pi}{2}}&e^{-i\frac{\pi}{4}}\\
      1&e^{+i\frac{3\pi}{4}}&e^{+i\frac{\pi}{2}}&e^{-i\frac{3\pi}{4}}\\
    \end{pmatrix},
\end{align*}
where the eigenvector $(1,1,1,1)^T$ corresponds to the $A_1$ representation 
($E=-3\gamma$), while the other three orthogonal eigenvectors 
correspond to the triply degenerate $T_2$ representation ($E=+\gamma$).

By discarding the lowest-energy flat band ($E=-3\gamma$), 
we obtain the following effective three-band Hamiltonian
around the $\Gamma_{\mathrm{high}}$ point:
\begin{align*}
    H_{\Gamma_{\mathrm{high}},\mathrm{eff}}(\bm{k}) =
    -\frac{\gamma a}{2}
    \begin{pmatrix}
      k_z&\frac{1}{\sqrt{2}}k_-&0\\
      \frac{1}{\sqrt{2}}k_+&0&\frac{1}{\sqrt{2}}k_-\\
      0&\frac{1}{\sqrt{2}}k_+&-k_z\\
    \end{pmatrix}.
\end{align*}
This can be written compactly as
\begin{align}
    H_{\Gamma_{\mathrm{high}},\mathrm{eff}}(\bm{k})
    = -\frac{\gamma a}{2} \bm{k} \cdot \bm{S}\,
    \label{eq:Gamma_eff_hamil}
\end{align}
where $\bm{S} = (S_x, S_y, S_z)$ denotes the spin-1 matrices:
\begin{widetext}
\begin{align*}
    \bm{S} = (S_x, S_y, S_z) 
    = \begin{pmatrix}
        \begin{pmatrix}
        0&\frac{1}{\sqrt{2}}&0\\
        \frac{1}{\sqrt{2}}&0&\frac{1}{\sqrt{2}}\\
        0&\frac{1}{\sqrt{2}}&0\
        \end{pmatrix},&
        \begin{pmatrix}
        0&-\frac{i}{\sqrt{2}}&0\\
        +\frac{i}{\sqrt{2}}&0&-\frac{i}{\sqrt{2}}\\
        1&+\frac{i}{\sqrt{2}}&0\
        \end{pmatrix},&
        \begin{pmatrix}
        1&0&0\\
        0&0&0\\
        0&0&-1\\
        \end{pmatrix}
    \end{pmatrix}.
\end{align*}

The corresponding eigenvalues are
\begin{align*}
    E_{\Gamma_{\mathrm{high}}}^- = -\frac{\gamma a}{2}k,\quad
    E_{\Gamma_{\mathrm{high}}}^0 = 0,\quad
    E_{\Gamma_{\mathrm{high}}}^+ = +\frac{\gamma a}{2}k, 
\end{align*}
corresponding to a lower conical band, 
a flat middle band, and an upper conical band,
meeting at the $\Gamma_{\mathrm{high}}$ point to form a \textit{triple Dirac cone}
[Fig.~\ref{fig:k4_Weyl_points}].

The normalized eigenfunctions of the effective 
Hamiltonian $H_{\Gamma_{\mathrm{high}},\mathrm{eff}}(\bm{k})$ 
[Eq.~\eqref{eq:Gamma_eff_hamil}] are given by
\begin{align*}
    \bm{u}_{\Gamma_{\mathrm{high}}}^- = \frac{1}{2k(k_z+k)}
    \begin{pmatrix}
    (k_z+k)^2 \\
    \sqrt2 k_+(k_z+k) \\
    k_+^2
    \end{pmatrix}, 
    \bm{u}_{\Gamma_{\mathrm{high}}}^0  = \frac{1}{\sqrt{2}k}
    \begin{pmatrix}
    -k_- \\
    \sqrt2 k_z \\
    k_+
    \end{pmatrix}, 
    \bm{u}_{\Gamma_{\mathrm{high}}}^+  = \frac{1}{2k(k_z-k)}
    \begin{pmatrix}
    (k_z-k)^2 \\
    \sqrt2 k_+(k_z-k) \\
    k_+^2
    \end{pmatrix}.
\end{align*}

The effective Hamiltonians at the $\mathrm{H_{low}}$, $\mathrm{P_{low}}$, 
and $\mathrm{P_{high}}$ points, which exhibit Weyl-type linear dispersions, 
are derived in Appendix~\ref{app:weyl_derivation}. 
Here, we summarize their explicit forms.

For the $\mathrm{H_{low}}$ point:
\begin{align*}
    H_{\mathrm{H_{low}},\mathrm{eff}}(\bm{k}) =
    +\frac{\gamma a}{2}
    \begin{pmatrix}
        k_z&\frac{1}{\sqrt{2}}k_-&0\\
        \frac{1}{\sqrt{2}}k_+&0&\frac{1}{\sqrt{2}}k_-\\
        0&\frac{1}{\sqrt{2}}k_+&-k_z\\
    \end{pmatrix}
    = +\frac{\gamma a}{2} \bm{k} \cdot \bm{S}.
\end{align*}
Note that the effective Hamiltonian at the $\mathrm{H_{low}}$ point 
has the same matrix structure as that at the $\Gamma_{\mathrm{high}}$ point,
but with an overall sign reversal in front of $\bm{k} \cdot \bm{S}$.

For the $\mathrm{P}$ point 
($\bm{k} = (\frac{\pi}{a}, \frac{\pi}{a}, \frac{\pi}{a})$),
two inequivalent Weyl points appear, 
denoted as $\mathrm{P_{low}}$ and $\mathrm{P_{high}}$, 
corresponding to distinct energy sectors 
$E = \mp \sqrt{3}\gamma$, respectively:
\begin{align*}
    H_{\mathrm{P_{low}},\mathrm{eff}}(\bm{k}) =
    +\frac{\gamma a}{6}
    \begin{pmatrix}
        k_x+k_y+k_z   &\sqrt{2}(\omega k_x+\omega^2 k_y+k_z)\\
        \sqrt{2}(\omega^2 k_x+\omega k_y+k_z)   &-k_x-k_y-k_z\\
    \end{pmatrix},
\end{align*}
and
\begin{align*}
    H_{\mathrm{P_{high}},\mathrm{eff}}(\bm{k}) =
    +\frac{\gamma a}{6}
    \begin{pmatrix}
        k_x+k_y+k_z   &\sqrt{2}(\omega^2 k_x+\omega k_y+k_z)\\
        \sqrt{2}(\omega k_x+\omega^2 k_y+k_z)   &-k_x-k_y-k_z\\
    \end{pmatrix},
\end{align*}
\end{widetext}
where $\omega = \frac{-1+i\sqrt{3}}{2},\quad \omega^3 = 1$.
The $\mathrm{P_{low}}$ and $\mathrm{P_{high}}$ Hamiltonians 
represent two symmetry-inequivalent linear band crossings 
originating from distinct doublet states in the tight-binding spectrum.

Since the effective Hamiltonian at the $\mathrm{P'}$ point 
is related to that at the $\mathrm{P}$ point by a unitary transformation, 
we restrict our discussion to the $\mathrm{P}$ point.

\subsection{Berry curvature and topological charge}
To evaluate the topological nature of each Weyl point, 
we compute the Berry connection and curvature from the corresponding
eigenfunctions.
Physically, the Berry connection can be regarded as a vector
potential in momentum space, and the Berry curvature as its associated
magnetic field in momentum space. 
The Berry connection $\bm{A}^j(\bm{k})$ for the $j$-th band is defined as
\begin{align*}
    \bm{A}^j(\bm{k}) &= (A^j_x(\bm{k}), A^j_y(\bm{k}), A^j_z(\bm{k})),
\end{align*}
where
\begin{align*}
    A^j_\alpha = i\langle{u}_j(\bm{k})|
    \frac{\partial}{\partial k_\alpha} |u_j(\bm{k})\rangle, \quad \alpha = x,y,z.
\end{align*} 
The Berry curvature $\bm{\Omega}_j(\bm{k})$ for $j$-th band is then given
as the rotation of Berry connection, i.e., 
\begin{align*}
    \bm{\Omega}_j(\bm{k}) &= \bm{\nabla_{k}} \times \bm{A}^j(\bm{k}),
\end{align*}
where $\bm{\nabla_{k}} = (\frac{\partial}{\partial k_x}, 
\frac{\partial}{\partial k_y}, \frac{\partial}{\partial k_z})$.

Weyl points are characterized by a topological charge $\phi$
~\cite{Chern1974, thouless1982quantized-6b7, elbistan2017weyl-ea9, Wan2011TSandFermiArcs}
and a chirality $\chi$
~\cite{weyl1929elektron-50c, Wan2011TSandFermiArcs},
which quantify their role as monopoles of the Berry curvature 
in momentum space. 
The topological charge $\phi_j$ of the $j$-th band 
is defined as the Berry flux through a closed surface $S$
enclosing the Weyl point,
\begin{align*}
    \phi_j = \int_S \bm{\Omega}_j \cdot \mathrm{d}\bm{S}. 
\end{align*}

The chirality $\chi$ of a Weyl point is given by the total Berry flux 
from all occupied bands, normalized by $2\pi$, 
when the Fermi level is located at the degeneracy:
\begin{align*}
    \chi = \frac{1}{2\pi} \sum_{j=1}^{N_\mathrm{occ.}} \phi_j. 
\end{align*}
A Weyl point with $\chi > 0$ acts as a source of Berry curvature, 
whereas $\chi < 0$ represents a sink.

\begin{figure}[h]
    \centering
    \includegraphics[width=0.8\linewidth]{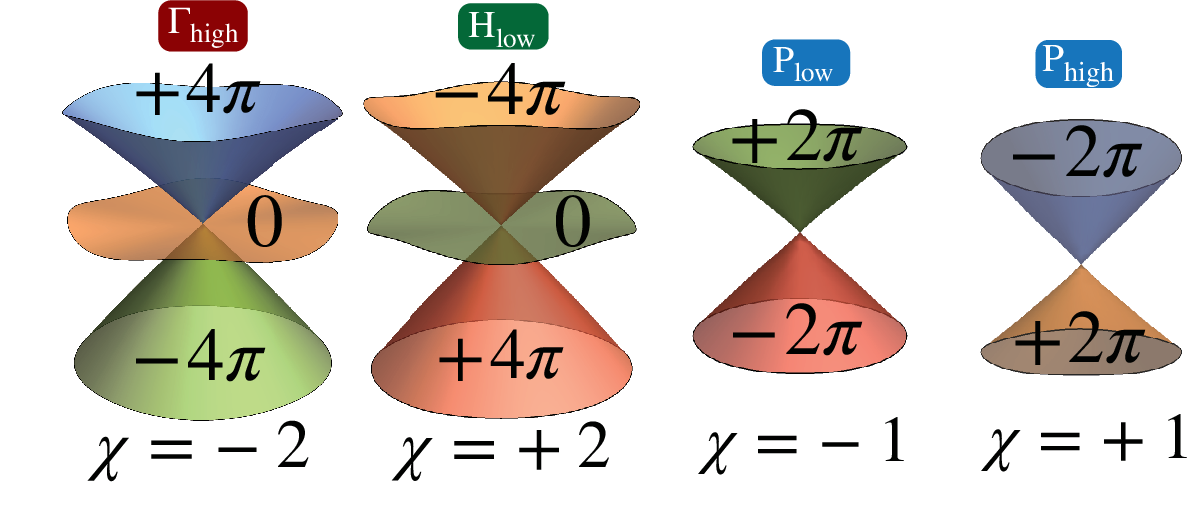}
    \caption{
        Energy dispersion around Weyl points in $K_4$ crystal.
        At the $\Gamma_{\mathrm{high}}$ and $\mathrm{H_{low}}$ points, 
        two conical dispersions 
        and one flat band intersect at a single point, 
        forming what is referred to as a \textbf{triple Dirac cone}.
        The topological charge is calculated for each band, 
        and the sign of the charge associated with the lowest band 
        determines the chirality of the degeneracy point.
        At the $\Gamma_{\mathrm{high}}$ point, the chirality is $\chi = -2$,
        while at the $\mathrm{H_{low}}$ point, it is $\chi = +2$.
        At the $\mathrm{P_{low}}$ and $\mathrm{P_{high}}$ point, 
        two conical bands intersect, forming what are referred to as
        \textbf{simple Dirac cones},
        each carrying opposite chirality:
        $\chi = -1$ at $\mathrm{P_{low}}$ 
        and $\chi = +1$ at $\mathrm{P_{high}}$.
    }\label{fig:k4_Weyl_points}
\end{figure}

The topological charges and chiralities of the Weyl points in the $K_4$ crystal, 
which characterize their monopole nature in momentum space, 
are summarized in Fig.~\ref{fig:k4_Weyl_points}.

At the $\Gamma_{\mathrm{high}}$ point, the Berry connections 
$\bm{A}_{\Gamma_{\mathrm{high}}}$ are obtained from the corresponding
eigenfunctions $\bm{u}_{\Gamma_{\mathrm{high}}}$ as
\begin{align*}
    \bm{A}_{\Gamma_{\mathrm{high}}}^- &= \frac{1}{k(k_z+k)}
    \begin{pmatrix}
    k_y\\-k_x\\0
    \end{pmatrix},\\
    \bm{A}_{\Gamma_{\mathrm{high}}}^0& = \bm{0},\\
    \bm{A}_{\Gamma_{\mathrm{high}}}^+& = \frac{1}{k(k_z-k)}
    \begin{pmatrix}
    -k_y\\k_x\\0
    \end{pmatrix}.
\end{align*}
The corresponding Berry curvatures are given by
\begin{align*}
    \bm{\Omega}_{\Gamma_{\mathrm{high}}}^- 
    =-\frac{\bm{k}}{k^3},\quad
    \bm{\Omega}_{\Gamma_{\mathrm{high}}}^0 = \bm{0},\quad 
    \bm{\Omega}_{\Gamma_{\mathrm{high}}}^+ 
    =+\frac{\bm{k}}{k^3}. 
\end{align*}
Thus, at the $\Gamma_{\mathrm{high}}$, the Berry curvature exhibits a
monopole-like distribution, isotropic in momentum space and decaying as
$1/k^2$ away from the degeneracy point.

At $\Gamma_{\mathrm{high}}$, the topological charges are higher in magnitude 
and remain quantized regardless of the choice of the enclosed surface $S$:
\begin{align*}
    \phi_{\Gamma_{\mathrm{high}}}^- 
    =-4\pi,\quad
    \phi_{\Gamma_{\mathrm{high}}}^0 = 0,\quad 
    \phi_{\Gamma_{\mathrm{high}}}^+ 
    =+4\pi. 
\end{align*}
The charge of the lower (upper) conical band is a higher topological charge 
(twice that of a conventional Weyl point), acting as a sink (source) of Berry curvature, 
while the crossing flat band carries zero charge.

The isolated band ($E = -3\gamma$) is topologically trivial, since its
Berry curvature and connection both vanish.
Therefore, the total chirality at $\Gamma_{\mathrm{high}}$ is
$\chi = -2$, given by the sum of $\phi_{\Gamma_{\mathrm{high}}}^-$ and
$\phi_{\Gamma_{\mathrm{high}}}^0$ divided by $2\pi$.

Topological properties of other degeneracy points, $\mathrm{H_{low}}$, 
$\mathrm{P_{low}}$, and $\mathrm{P_{high}}$ are analytically derived
in Appendix~\ref{app:weyl_derivation}.

For the $\mathrm{H_{low}}$ point, the Berry curvatures are
\begin{align*}
    \bm{\Omega}_{\mathrm{H_{low}}}^- = 
    +\frac{\bm{k}}{k^3},\quad
    \bm{\Omega}_{\mathrm{H_{low}}}^0 &= \bm{0},\quad
    \bm{\Omega}_{\mathrm{H_{low}}}^+ = 
    -\frac{\bm{k}}{k^3}.
\end{align*}
This has the same form as that at $\Gamma_{\mathrm{high}}$,
but with opposite sign.
Consequently, the $\mathrm{H_{low}}$ point possesses 
a total chirality of $\chi = +2$, indicating that it is 
a Weyl node with higher topological charge, opposite in sign to $\Gamma_{\mathrm{high}}$.

For the $\mathrm{P_{low}}$ and $\mathrm{P_{high}}$ points, the Berry
curvatures are obtained as
\begin{align*}
    \bm{\Omega}_\mathrm{P_{low}}^{-} = 
    \frac{-\bm{k}}{2k^3}, \quad
    \bm{\Omega}_\mathrm{P_{low}}^{+}\ = 
    \frac{+\bm{k}}{2k^3},
\end{align*}
and 
\begin{align*}
    \bm{\Omega}_\mathrm{P_{high}}^{-} = 
    \frac{+\bm{k}}{2k^3}, \quad
    \bm{\Omega}_\mathrm{P_{high}}^{+} = 
    \frac{-\bm{k}}{2k^3}. 
\end{align*}
The signs of the upper and lower bands, as well as those of
$\mathrm{P_{low}}$ and $\mathrm{P_{high}}$, are reversed.

These Berry curvatures have half the amplitude compared to 
those at $\Gamma_{\mathrm{high}}$ and $\mathrm{H_{low}}$, 
indicating that $\mathrm{P_{low}}$ and $\mathrm{P_{high}}$ 
correspond to conventional Weyl points.

Thus, at all Weyl points in the $K_4$ crystal
($\Gamma_{\mathrm{high}}$, $\mathrm{H_{low}}$, $\mathrm{P_{low}}$, and
$\mathrm{P_{high}}$), the Berry curvature exhibits a monopole-like distribution.
According to the Nielsen-Ninomiya theorem
~\cite{NielsenNinomiya1,NielsenNinomiya2,Armitage2018WeylandDirac}, 
the total chirality over the entire Brillouin zone (BZ) must vanish.
Therefore, Weyl points always appear in pairs or sets whose chiralities sum to zero.
In the $K_4$ crystal, this condition is satisfied, 
as the BZ contains one symmetry-equivalent $\Gamma$ (or H) point 
and two symmetry-equivalent P points, whose total chirality exactly cancels to zero.

In systems preserving time-reversal symmetry but lacking inversion symmetry,
Weyl points with the same chirality must occur at momenta $\bm{k}$ and $-\bm{k}$,
since
\begin{align*}
    \bm{\Omega}(\bm{k}) = -\bm{\Omega}(-\bm{k})
    \quad \Rightarrow \quad
    \chi(\bm{k}) = \chi(-\bm{k}).
\end{align*}

A distinctive feature of the $K_4$ crystal is the existence of a Weyl
point at $\Gamma_{\mathrm{high}}$ that lacks a counterpart at the
opposite momentum.
In systems with time-reversal symmetry but without inversion symmetry,
Weyl points must appear in pairs with the same chirality at $\bm{k}$ and
$-\bm{k}$, owing to the relation
\begin{align*}
    \bm{\Omega}(\bm{k}) = -\bm{\Omega}(-\bm{k})
    \quad \Rightarrow \quad
    \chi(\bm{k}) = \chi(-\bm{k}).
\end{align*}
In the $K_4$ crystal, the H and P points satisfy this condition due to
their positions on the Brillouin zone (BZ) boundary, as they are mapped onto
themselves or symmetry-equivalent points under momentum inversion.
In contrast, the $\Gamma$ point lies at the BZ center and effectively
pairs with itself.

Because the Weyl points appear at different energies, 
tuning the Fermi level near a single Weyl point allows the system to 
exhibit its associated topological effects independently—such as 
chiral-anomaly-induced transport or asymmetric surface states-without interference 
from other Weyl nodes.

It is also worth noting that recent works have explored 
the role of crystal symmetry - especially screw and rotational
symmetries - 
in realizing Weyl nodes with higher chiral charges (for example, 
Ref.~\cite{tsirkin2017composite-e05}, and Ref.~\cite{gonzalezhernandez2021chiralities-0b4}). 
While such symmetry-enforced, high-chirality nodes lie beyond the scope 
of the present analysis, they suggest promising avenues for future investigation.

\section{Fermi Arc}
In $K_4$ crystal, the presence of Weyl points gives rise to 
topological surface states-so-called Fermi arcs-that appear 
when we consider a system with surfaces. 
We first briefly review the origin of Fermi arcs in conventional Weyl semimetals, 
and then discuss the characteristic features of those in the $K_4$ crystal.

\subsection{Conventional Weyl semimetals and Fermi arcs}
In conventional Weyl semimetals, pairs of Weyl points with opposite chirality 
are connected by surface states known as Fermi arcs
~\cite{Wan2011TSandFermiArcs, Armitage2018WeylandDirac}. 
Their topological origin can be understood by considering 
a two-dimensional cylindrical subsystem in momentum space. 
The cylinder is oriented perpendicular to the slab Brillouin zone (BZ), 
for example along $k_z$ in a (001) slab geometry, 
and encloses a single Weyl point in the bulk BZ. 
On this cylindrical manifold, the energy spectrum is fully gapped 
except at the enclosed Weyl point. 
The Chern number~\cite{thouless1982quantized-6b7, kohmoto1993berrys-d35} 
defined on the cylinder equals the chirality of the Weyl point, 
provided that no other nodes lie within it. 
If multiple Weyl points are enclosed, 
the Chern number reflects the sum of their chiralities. 
When this nonzero Chern number is projected onto the slab BZ, 
the resulting topologically protected edge states
~\cite{Kane2005, hatsugai1993chern-a9b} manifest as Fermi arcs 
that terminate at the projections of Weyl points with opposite chirality.

\subsection{Fermi arcs in the $K_4$ crystal}
\begin{figure*}
\centering
\includegraphics[width=1\linewidth]{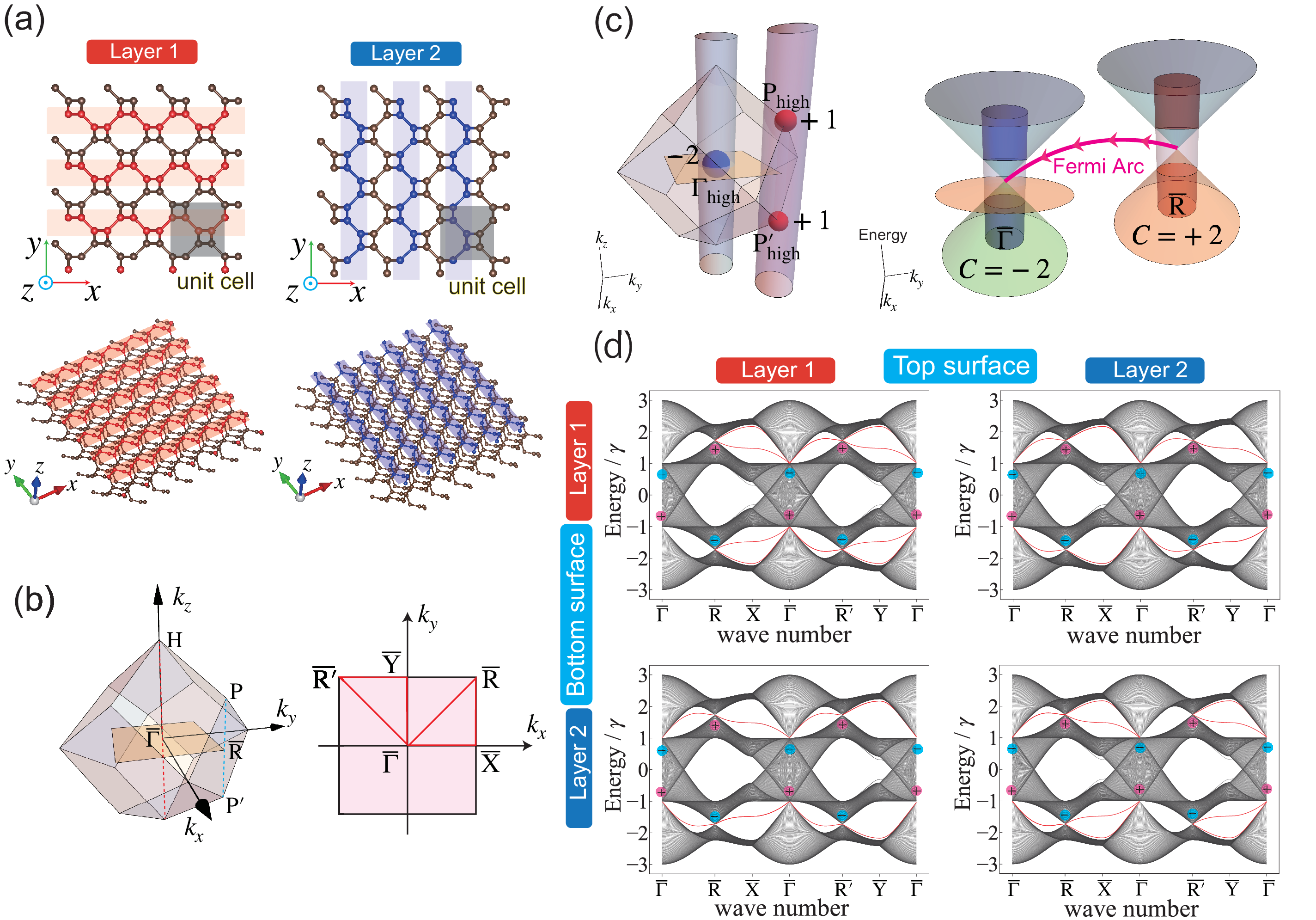}
\caption{
(a) Slab structure of the $K_4$ crystal cleaved along (001). 
Two different types of surface structures: (left) Layer 1 (red) forms
 chains along the $x$-axis, and (right) Layer 2 (blue) along the $y$-axis.  
Top and perspective views are shown in the upper and lower panels, respectively. 
Thus, for the slab structures, there are totally four combinations of
 possible surface terminations. 
(b) 1st BZ for Bulk and slab strcutures.
(left) 1st BZ of bulk structure, with the yellow square indicating 
the slab BZ obtained by projecting along the $k_z$-direction. 
(right) 1st BZ of slab structures.
$\overline{\Gamma}$ corresponds to the projection of the bulk $\Gamma$
 and H points, while $\overline{\mathrm{R}}$  
correspond to projections of the bulk P and P' points. 
(c) 
Left panel schematically shows the bulk BZ, 
where $\Gamma_{\mathrm{high}}$ (blue sphere) carries chirality $\chi = -2$, 
and $\mathrm{P_{high}}$ and $\mathrm{P'_{high}}$ (red spheres) each carry $\chi = +1$. 
These Weyl nodes are enclosed by distinct momentum-space subsystems: 
$\Gamma_{\mathrm{high}}$ by a cylinder centered at $\overline{\Gamma}$ (blue), 
and $\mathrm{P_{high}}$ and $\mathrm{P'_{high}}$ by a cylinder centered at $\overline{\mathrm{R}}$ (red). 
Accordingly, the subsystem centered at $\overline{\Gamma}$ acquires a Chern number $C = -2$, 
while that centered at $\overline{\mathrm{R}}$ acquires $C = +2$. 
Right panel shows the energy band structures of slab at $\overline{\Gamma}$ 
and $\overline{\mathrm{R}}$, respectively.
In both cases, the subsystems cut through Weyl cones, yielding effectively two-dimensional systems 
with energy gaps that host nontrivial Chern numbers. 
Consequently, topologically protected edge states traverse these gaps. 
Since the bulk gap closes only at the Weyl nodes, 
the surface states connect their projections, 
forming a Fermi arc (pink line) that extends from 
$\overline{\mathrm{R}}$ ($C = +2$) to $\overline{\Gamma}$ ($C = -2$).
(d) Energy dispersion of slab structures obtained from the tight-binding model. 
Gray shading indicates the bulk projection, and red curves highlight the Fermi arcs. 
Weyl points are labeled by their chiralities. 
High-energy arcs connect the projected $\Gamma_{\mathrm{high}}$ ($\chi = -2$) 
to $\mathrm{P_{high}}$ and $\mathrm{P'_{high}}$ ($\chi = +1 \times 2$), 
while low-energy arcs connect the projected $\mathrm{H_{low}}$ ($\chi = +2$) 
to $\mathrm{P_{low}}$ and $\mathrm{P'_{low}}$ ($\chi = -1 \times 2$). 
The apparent chirality imbalance is resolved by accounting for 
the projection of symmetry-equivalent Weyl points in the bulk. }
\label{fig:k4slab}
\end{figure*}

We next examine the Fermi arcs that emerge in the $K_4$ crystal, 
which connect Weyl points carrying large topological charges.
To analyze these surface states, 
we construct a slab geometry by cleaving the bulk crystal 
along the $xy$ plane, as illustrated in Fig.~\ref{fig:k4slab}(a).
The three-dimensional bulk BZ is projected onto a two-dimensional slab BZ,
as shown in Fig.~\ref{fig:k4slab}(b).
The projections of bulk Weyl points onto the slab BZ 
are summarized in Table~\ref{tab:bulk-slab}.
The $\Gamma_{\mathrm{high}}$ and $\mathrm{H_{low}}$ points project 
onto the surface $\mathrm{\overline{\Gamma}}$ point, 
while $\mathrm{P_{low}}$ and $\mathrm{P_{high}}$ 
project onto $\mathrm{\overline{R}}$.
As shown in Fig.~\ref{fig:k4slab}(c), the bulk gap between the upper two bands
closes at the $\Gamma_{\mathrm{high}}$, $\mathrm{P_{high}}$, 
and $\mathrm{P'_{high}}$ points. 
Consequently, Fermi arcs emerge across this energy range in the slab system.

The subsystem surrounding the $\overline{\Gamma}$ point reflects 
the chirality projected from the bulk $\Gamma_{\mathrm{high}}$ point. 
Owing to the presence of a triple Dirac cone at $\Gamma_{\mathrm{high}}$ 
with chirality $\chi = -2$, the resulting 2D subsystem acquires 
a higher Chern number of $C = -2$. 
On the other hand, the subsystem centered at the $\overline{\mathrm{R}}$ point 
encloses the projected Weyl points at $\mathrm{P_{high}}$ and $\mathrm{P'_{high}}$, 
each with chirality $\chi = +1$, yielding a total Chern number $C = +2$ for this subsystem. 
These nonzero Chern numbers ensure the existence of Fermi arc surface states 
crossing the gap, connecting $\overline{\Gamma}$ and $\overline{\mathrm{R}}$ points 
via projected Weyl nodes.

Similarly, the gap between the lower two bands closes at $\mathrm{H_{low}}$, 
$\mathrm{P_{low}}$, and $\mathrm{P'_{low}}$ points. 
Here, $\mathrm{H_{low}}$ is projected onto $\overline{\Gamma}$, 
forming a subsystem with $C = +2$, 
while $\mathrm{P_{low}}$ and $\mathrm{P'_{low}}$ project onto $\overline{\mathrm{R}}$, 
contributing $C = -2$ in total. 
Fermi arcs also appear between these projections in the lower energy sector.

These features were confirmed by calculating the slab band structures 
using the tight-binding model. As shown in Fig.\ref{fig:k4slab}(a), 
the (001) slab consists of alternating layers, allowing for four different slab terminations 
depending on which layer appears on each surface. 
Figure~\ref{fig:k4slab}(d) displays the band structures of all four terminations. 
In each case, Fermi arcs appear in both the high-energy region ($E > 0$), 
connecting $\Gamma_{\mathrm{high}}$ to $\mathrm{P_{high}}$ and $\mathrm{P'_{high}}$, 
and the low-energy region ($E < 0$), connecting $\mathrm{H_{low}}$ 
to $\mathrm{P_{low}}$ and $\mathrm{P'}_{\mathrm{low}}$.

A key distinction from conventional Weyl systems is that 
one end of a Fermi arc attaches to a triple Dirac cone 
with chirality $\chi = \pm 2$, 
while the other end splits into two projected Weyl nodes with $\chi = \mp 1$, 
preserving overall charge neutrality. 
Thus, a single arc pair involves three Weyl nodes at different energies. 
Because these arcs are topologically protected, 
they persist even when Weyl nodes acquire small gaps, 
providing a route to higher-mobility surface states 
and potential device applications based on higher Chern numbers $C = \pm
2$.

\begin{table}[h]
    \caption{Weyl points (bulk) and their projections onto slab BZ}
    \label{tab:bulk-slab}
    \centering
    \begin{tabular}{l c c c c}
        \hline\hline
        Weyl Point (bulk) & $\Gamma_{\mathrm{high}}$ & $\mathrm{H_{low}}$ 
        & $\mathrm{P_{low}}$ & $\mathrm{P_{high}}$ \\
        \hline
        Projected Point in Slab BZ 
        & $\overline{\Gamma}$ & $\overline{\Gamma}$ & $\mathrm{\overline{R}}$ & $\mathrm{\overline{R}}$ \\
        Energy  
        & $+\gamma$ & $-\gamma$ & $-\sqrt{3}\gamma$ & $+\sqrt{3}\gamma$ \\
        Chirality $\chi$
        & -2 & +2 & -1 & +1 \\
        Symmetry Multiplicity
        & 1 & 1 & 2 & 2 \\
        \hline \hline
    \end{tabular}
\end{table}

\section{Summary}\label{sec:summary}
We have investigated the topological electronic properties of the $K_4$ crystal,
a three-dimensional analogue of graphene, by constructing a tight-binding model
and analyzing its band structure.
The bulk Brillouin zone (BZ) hosts triple band degeneracies at
$\Gamma_{\mathrm{high}}$ and $\mathrm{H_{low}}$,
comprising two linearly dispersing bands and one flat band-forming
what is known as a triple Dirac cone.
In contrast, double degeneracies occur at $\mathrm{P_{low}}$ and $\mathrm{P_{high}}$,
resulting in conventional Dirac cones.

These degeneracy points act as Weyl nodes characterized by isotropic Berry curvature
and quantized topological charge (chirality):
$\chi = -2$ and $+2$ at $\Gamma_{\mathrm{high}}$ and $\mathrm{H_{low}}$,
and $\chi = -1$ and $+1$ at $\mathrm{P_{low}}$ and $\mathrm{P_{high}}$, respectively.

To examine the associated surface states, we constructed a slab geometry
by cleaving the bulk crystal along the $(001)$ direction.
In this configuration, the Weyl nodes at $\Gamma_{\mathrm{high}}$ and $\mathrm{H_{low}}$
project onto the surface $\overline{\Gamma}$ point,
while those at $\mathrm{P_{low}}$ and $\mathrm{P_{high}}$ project onto $\overline{R}$.
Although the Fermi arcs appear to connect Weyl points with unbalanced chiralities,
this apparent asymmetry is resolved by the presence of two symmetry-equivalent
$\mathrm{P_{low}}$ or $\mathrm{P_{high}}$ points,
which effectively double the projected chirality at $\overline{R}$.
This restores the topological charge balance between $\overline{\Gamma}$ and $\overline{R}$.

Consequently, surface states in the form of Fermi arcs emerge,
linking the triple Dirac cone to pairs of simple Dirac cones
in a topologically consistent manner.
Our results establish the $K_4$ crystal as a novel Weyl semimetal
where higher-chirality nodes are compensated by multiple lower-chirality ones,
with their projections robustly connected by topologically protected surface states.
This work highlights the $K_4$ crystal as an archetype of
three-dimensional $sp^2$ carbon networks with intrinsic topological properties,
providing a foundation for future exploration of higher-chirality fermions
and their potential applications in topological electronics.

Further experimental validation of the predicted topological phenomena 
in the $K_4$ crystal will be essential. 
For instance, angle-resolved photoemission spectroscopy (ARPES) and 
quantum-oscillation measurements have already observed 
Fermi arcs and Berry phase signatures in Weyl semimetals 
such as TaAs (Refs.~\cite{lv2015experimental-9e6, belopolski2016criteria-7fa, 
huang2015observation-437, hu2016pi-10f}) 
and large-Chern-number systems like PtGa (Ref.~\cite{yao2020observation-d2f}). 
Drawing on these precedents, similar techniques could be applied 
once a realisation of the $K_4$ crystal becomes available. 
Additionally, photonic-crystal analogues of the triple-degenerate point 
have recently been demonstrated in acoustic/photonic systems 
(for example, Refs.~\cite{yang2019topological-919, lu2013weyl-f80}), 
suggesting that the $K_4$ lattice-with its inherent higher-chirality fermions
could serve as a prototype not only for electronic topological materials 
but also for topological photonics and metamaterials.

Looking ahead, an intriguing direction is to determine whether the
chirality and the Berry curvature of the nodal points can be derived
purely from crystal symmetry, for example through symmetry indicators or
irreducible-representation analysis. 
Establishing such a connection would not only reinforce the topological
robustness of the observed nodes, but also enable a deeper
symmetry-based classification of topological semimetals in nonsymmorphic
crystals such as the $K_4$ lattice.

\section*{ACKNOWLEDGMENTS}
This work was supported by JSPS KAKENHI (Grants No. JP25K01609,
No. JP22H05473, and No. JP21H01019), JST CREST (Grant
No. JPMJCR19T1). K.W. acknowledges the financial support for Basic
Science Research Projects (Grant No. 2401203) from the Sumitomo
Foundation.  

\section*{DATA AVAILABILITY}
The data used and analyzed during the current study available from corresponding authors on reasonable request.

\newpage
\appendix
\begin{widetext}
\section{Topological properties at $\mathrm{H_{low}}$,
 $\mathrm{P_{low}}$, $\mathrm{P_{high}}$ points}
\label{app:weyl_derivation}
In this section, we provide the derivations of the topological
quantities such as Berry connection, Berry curvature and topological
charges
at $\mathrm{H_{low}}$, $\mathrm{P_{high}}$, and $\mathrm{P_{low}}$ points.

\subsection{$\mathrm{H_{low}}$ point}
Expanding the tight-binding Hamiltonian [Eq.~\eqref{eq:k4hamil}] 
to first order in momentum $\bm{k}$ around the $\mathrm{H}$ point,
i.e., $\bm{k}_\mathrm{H} = (0,\, \frac{2\pi}{a},\, 0)$, 
the effective Hamiltonian can be written as
\begin{align*}
    H_\mathrm{H}^{(1)}(\bm{k}) 
    &= -\gamma
    \begin{pmatrix}
        0&-i&1&i\\
        i&0&-i&1\\
        1&i&0&-i\\
        -i&1&i&0
    \end{pmatrix}
    -\frac{\gamma a}{4}
    \begin{pmatrix}
        0&-(k_y-k_z)&-i(k_z-k_x)&(k_x-k_y) \\
        -(k_y-k_z)&0&-(k_x+k_y)&-i(-k_z-k_x) \\
        i(k_z-k_x)&-(k_x+k_y)&0&-(k_y+k_z) \\
        (k_x-k_y)&i(-k_z-k_x)&-(k_y+k_z)&0 \\
    \end{pmatrix}.
\end{align*}
This Hamiltonian can be block-diagonalized using 
a unitary matrix $U_{\mathrm{H}}$, as
\begin{align*}
  U_\mathrm{H}^\dagger H_\mathrm{H}^{(1)}(\bm{k}) U_\mathrm{H} &=
  -\gamma
  \begin{pmatrix}
    -3&0&0&0\\
    0&1&0&0\\
    0&0&1&0\\
    0&0&0&1\\
  \end{pmatrix}
  +\frac{\gamma a}{2}
  \begin{pmatrix}
    0&0&0&0\\
    0&k_z&\frac{1}{\sqrt{2}}k_-&0\\
    0&\frac{1}{\sqrt{2}}k_+&0&\frac{1}{\sqrt{2}}k_-\\
    0&0&\frac{1}{\sqrt{2}}k_+&-k_z\\
  \end{pmatrix},
\end{align*}
where $k = \sqrt{k_x^2+k_y^2+k_z^2}$ and $k_{\pm} = k_x\pm ik_y$. 
\end{widetext}
The unitary matrix $U_\mathrm{H}$ can be given as 
\begin{align*}
    U_\mathrm{\mathrm{H}} = \frac{1}{2}
    \begin{pmatrix}
        -i&e^{-i\frac{3\pi}{4}}&1&e^{-i\frac{\pi}{4}}\\
        -1&e^{i\frac{\pi}{4}}&i&e^{-i\frac{\pi}{4}}\\
        i&e^{i\frac{3\pi}{4}}&1&e^{i\frac{\pi}{4}}\\
        1&e^{i\frac{3\pi}{4}}&i&e^{-i\frac{3\pi}{4}}\\
    \end{pmatrix}.
\end{align*}

By discarding the term concerning the flat band ($+3\gamma$), 
we obtain an effective three-band Hamiltonian 
near the $\mathrm{H_{low}}$ point:
\begin{align*}
    H_{\mathrm{H_{low}},\mathrm{eff}}(\bm{k}) =
    +\frac{\gamma a}{2}
    \begin{pmatrix}
        k_z&\frac{1}{\sqrt{2}}k_-&0\\
        \frac{1}{\sqrt{2}}k_+&0&\frac{1}{\sqrt{2}}k_-\\
        0&\frac{1}{\sqrt{2}}k_+&-k_z\\
    \end{pmatrix}.
\end{align*}
This effective Hamiltonian can be expressed in a compact form as
\begin{align}
    H_{\mathrm{H_{low}},\mathrm{eff}}(\bm{k}) 
    = +\frac{\gamma a}{2} \bm{k} \cdot \bm{S}.
    \label{eq:H_eff_hamil}
\end{align}
This Hamiltonian is the sign-reversed counterpart of 
the effective Hamiltonian at the $\Gamma_{\mathrm{high}}$ point 
[Eq.~\eqref{eq:Gamma_eff_hamil}]. 
Its eigenvalues are
\begin{align*}
    E_{\mathrm{H_{low}}}^- &= \frac{-\gamma a}{2}k ,&
    E_{\mathrm{H_{low}}}^0 &= 0 ,&
    E_{\mathrm{H_{low}}}^+ &= \frac{+\gamma a}{2}k ,\\
\end{align*}
which are identical to those at the $\Gamma_{\mathrm{high}}$ point.

These correspond to a lower conical band, a flat central band, 
and an upper conical band, which meet at the $\mathrm{H_{low}}$ point 
to form a \textit{triple Dirac cone}, 
as illustrated in Fig.~\ref{fig:k4_Weyl_points} (the $\mathrm{H_{low}}$ point). 

However, the eigenspaces corresponding to the upper and lower conical bands 
at $\mathrm{H_{low}}$ are reversed compared to those at $\Gamma_{\mathrm{high}}$. 
The normalized eigenfunctions are given by
\begin{align*}
    \bm{u}_{\mathrm{H_{low}}}^- &= \frac{1}{2k(k_z-k)}
    \begin{pmatrix}
    (k_z-k)^2 \\
    \sqrt2 k_+(k_z-k) \\
    k_+^2
    \end{pmatrix},
    \\
    \bm{u}_{\mathrm{H_{low}}}^0 &= \frac{1}{\sqrt{2}k}
    \begin{pmatrix}
    -k_- \\
    \sqrt2 k_z \\
    k_+
    \end{pmatrix},
    \\
    \bm{u}_{\mathrm{H_{low}}}^+ &= \frac{1}{2k(k_z+k)}
    \begin{pmatrix}
    (k_z+k)^2 \\
    \sqrt2 k_+(k_z+k) \\
    k_+^2
    \end{pmatrix}.\\
\end{align*}
As a result, the Berry connections and Berry curvatures 
for these bands are also interchanged.

The Berry connections are
\begin{align*}
    \bm{A}_{\mathrm{H_{low}}}^- &= \frac{1}{k(k_z-k)}
    \begin{pmatrix}
    -k_y\\k_x\\0
    \end{pmatrix},
    \\
    \bm{A}_{\mathrm{H_{low}}}^0 &= \bm{0},
    \\
    \bm{A}_{\mathrm{H_{low}}}^+ &= \frac{1}{k(k_z+k)}
    \begin{pmatrix}
    k_y\\-k_x\\0
    \end{pmatrix},\\
\end{align*}
and the corresponding Berry curvatures are
\begin{align*}
    \bm{\Omega}_{\mathrm{H_{low}}}^- &
    =\frac{+\bm{k}}{k^3},
    &
    \bm{\Omega}_{\mathrm{H_{low}}}^0 &= \bm{0},
    &
    \bm{\Omega}_{\mathrm{H_{low}}}^+ &
    =\frac{-\bm{k}}{k^3}.\\
\end{align*}

The monopole charges associated with each band are then given by
\begin{align*}
    \phi_{\mathrm{H_{low}}}^- &= +4\pi,&
    \phi_{\mathrm{H_{low}}}^0 &= 0,&
    \phi_{\mathrm{H_{low}}}^+ &= -4\pi.
\end{align*}

Thus, the $\Gamma_{\mathrm{high}}$ and $\mathrm{H_{low}}$ points 
not only share the same band structure form but also 
exhibit opposite topological characteristics, 
forming a pair in both energy and topology.

\subsection{$\mathrm{P_{low}}$ point and $\mathrm{P_{high}}$ point}
The Hamiltonian matrix [Eq.~\eqref{eq:k4hamil}] 
is expanded up to first order in momentum 
$\bm{k}$ around the P point,
$\bm{k}_\mathrm{P} = (\frac{\pi}{a} ,\; \frac{\pi}{a} ,\; \frac{\pi}{a})$, as 

\begin{widetext}
\begin{align*}
    H_\mathrm{P}^{(1)}(\bm{k}) 
    &= -\gamma
    \begin{pmatrix}
        0&1&1&1 \\
        1&0&-i&i \\
        1&i&0&-i \\
        1&-i&i&0 \\
    \end{pmatrix}
    -\frac{\gamma a}{4}
    \begin{pmatrix}
        0&-i(k_y-k_z)&-i(k_z-k_x)&-i(k_x-k_y) \\
        i(k_y-k_z)&0&-(k_x+k_y)&-(-k_z-k_x) \\
        i(k_z-k_x)&-(k_x+k_y)&0&-(k_y+k_z) \\
        i(k_x-k_y)&-(-k_z-k_x)&-(k_y+k_z)&0 \\
    \end{pmatrix}.
\end{align*}
\end{widetext}

Although this Hamiltonian cannot be fully block-diagonalized 
by a unitary transformation, we consider the unitary matrix
\begin{align*}
    U_\mathrm{\mathrm{P}} = \frac{1}{\sqrt{6}}
    \begin{pmatrix}
        \sqrt{3}&0&-\sqrt{3}&0\\
        1&\sqrt{2}\omega&1&\sqrt{2}\omega^2\\
        1&\sqrt{2}\omega^2&1&\sqrt{2}\omega\\
        1&\sqrt{2}&1&\sqrt{2}\\
    \end{pmatrix},
\end{align*}
where $\omega = \frac{-1+i\sqrt{3}}{2},\quad \omega^3 = 1$.
This unitary matrix diagonalizes the zeroth-order part of 
$H_\mathrm{P}^{(1)}(\bm{k})$, 
i.e., the Hamiltonian at the P point itself.

By applying this transformation, we obtain
\begin{align*}
    U_\mathrm{P}^\dagger H_\mathrm{P}^{(1)}(\bm{k}) U_\mathrm{P} &=
    \begin{pmatrix}
        -\sqrt{3}\gamma&0&0&0\\
        0&-\sqrt{3}\gamma&0&0\\
        0&0&+\sqrt{3}\gamma&0\\
        0&0&0&+\sqrt{3}\gamma\\
    \end{pmatrix}
    +\delta H_{\mathrm{P}}(\bm{k}),
\end{align*}
where $\delta H_{\mathrm{P}}(\bm{k})$ contains all the first-order terms:
\begin{widetext}
\begin{align*}
    \delta H_{\mathrm{P}}(\bm{k}) =
    \begin{pmatrix}
        k_x+k_y+k_z & \sqrt{2}(\omega k_x+\omega^2 k_y+k_z) & k_x+k_y+k_z & \frac{1}{\sqrt{2}}(-\omega^2 k_x-\omega k_y-k_z)\\
        \sqrt{2}(\omega^2k_x+\omega k_y+k_z) & -k_x-k_y-k_z & \frac{1}{\sqrt{2}}(-\omega^2 k_x-\omega k_y-k_z) & -\omega k_x-\omega^2 k_y-k_z\\
        k_x+k_y+k_z &  \frac{1}{\sqrt{2}}(-\omega k_x-\omega^2 k_y-k_z) & k_x+k_y+k_z & \sqrt{2}(\omega^2 k_x+\omega k_y+k_z)\\
        \frac{1}{\sqrt{2}}(-\omega k_x-\omega^2 k_y-k_z) & -\omega^2 k_x-\omega k_y-k_z & \sqrt{2}(\omega k_x+\omega^2 k_y+k_z) & -k_x-k_y-k_z
    \end{pmatrix},
\end{align*}
\end{widetext}
which are not block-diagonalized by $U_\mathrm{P}$.

In contrast to the cases at the $\Gamma$ and $\mathrm{H}$ points, 
the first-order correction $\delta H_{\mathrm{P}}(\bm{k})$ 
is not diagonal in this transformed basis. 
However, since the zeroth-order Hamiltonian has a two-fold degenerate
low-energy sector ($-\sqrt{3}\gamma$) and 
high-energy sector ($+\sqrt{3}\gamma$), 
and the energy splitting is finite at $\bm{k} = 0$, 
we may treat these sectors separately in the vicinity of the P point.

Thus, by neglecting the off-diagonal coupling between 
the low-energy and high-energy subspaces in $\delta H_{\mathrm{P}}(\bm{k})$, 
we obtain two effective $2 \times 2$ Hamiltonians. 
The upper-left $2 \times 2$ block corresponds to 
the low-energy effective Hamiltonian at the $\mathrm{P_{low}}$ point, 
and the lower-right $2 \times 2$ block corresponds to 
the high-energy effective Hamiltonian at the $\mathrm{P_{high}}$ point.

The effective Hamiltonians are given by:
\begin{widetext}
\begin{align*}
    H_{\mathrm{P_{low}},\mathrm{eff}}(\bm{k}) &=
    +\frac{\gamma a}{6}
    \begin{pmatrix}
        k_x+k_y+k_z   &\sqrt{2}(\omega k_x+\omega^2 k_y+k_z)\\
        \sqrt{2}(\omega^2 k_x+\omega k_y+k_z)   &-k_x-k_y-k_z\\
    \end{pmatrix},
\\
    H_{\mathrm{P_{high}},\mathrm{eff}}(\bm{k}) &=
    +\frac{\gamma a}{6}
    \begin{pmatrix}
        k_x+k_y+k_z   &\sqrt{2}(\omega^2 k_x+\omega k_y+k_z)\\
        \sqrt{2}(\omega k_x+\omega^2 k_y+k_z)   &-k_x-k_y-k_z\\
    \end{pmatrix}.
\end{align*}
\end{widetext}

The eigenvalues of the effective Hamiltonian 
$H_{\mathrm{P_{low}},\mathrm{eff}}(\bm{k})$ are given by
\begin{align*}
    E_\mathrm{P_{low}}^{\pm} &= \pm \frac{\gamma a}{2\sqrt{3}}k,
\end{align*}
which describe a pair of conically dispersing bands meeting at 
the $\mathrm{P_{low}}$ point, thereby forming a simple Dirac cone, 
as shown in Fig.~\ref{fig:k4_Weyl_points} ($\mathrm{P_{low}}$).

The corresponding normalized eigenvectors are
\begin{align*}
    \bm{u}_\mathrm{P_{low}}^{-} &= 
\frac{1}{\eta_-}
    \begin{pmatrix}
    \frac{1}{\sqrt{2}}(k_x+k_y+k_z-\sqrt{3}k) \\
    \omega^2k_x+\omega k_y +k_z
    \end{pmatrix},
    \\
    \bm{u}_\mathrm{P_{low}}^{+} &= 
\frac{1}{\eta_+}
    \begin{pmatrix}
    \frac{1}{\sqrt{2}}(k_x+k_y+k_z+\sqrt{3}k) \\
    \omega^2k_x+\omega k_y +k_z
    \end{pmatrix},
\end{align*}
where $\eta_\pm ={\sqrt{\sqrt{3}k(\sqrt{3}k\pm k_x\pm k_y\pm k_z)}}$.

From these eigenstates, the Berry connections are computed as
\begin{align*}
    \bm{A}_\mathrm{P_{low}}^{-} &= \frac{1}{2k(k_x+k_y+k_z-\sqrt{3}k)}
    \begin{pmatrix}
    k_y-k_z\\k_z-k_x\\k_x-k_y
    \end{pmatrix},
    \\
    \bm{A}_\mathrm{P_{low}}^{+} &= \frac{1}{2k(k_x+k_y+k_z+\sqrt{3}k)}
    \begin{pmatrix}
    -k_y+k_z\\-k_z+k_x\\-k_x+k_y
    \end{pmatrix}.\\
\end{align*}
The corresponding Berry curvatures take the form
\begin{align*}
    \bm{\Omega}_\mathrm{P_{low}}^{-} 
    &=\frac{-\bm{k}}{2k^3},
    &
    \bm{\Omega}_\mathrm{P_{low}}^{+} 
    &=\frac{+\bm{k}}{2k^3}.\\
\end{align*}
These expressions indicate that the $\mathrm{P_{low}}$ point carries 
monopole-type Berry curvature, with magnitude equal to 
half of those found at the $\mathrm{\Gamma_{high}}$ and $\mathrm{H_{low}}$ 
points for the upper and lower cones.

Accordingly, the topological charges are given by
\begin{align*}
    \phi_\mathrm{P_{low}}^{-} &= -2\pi,&
    \phi_\mathrm{P_{low}}^{+} &= +2\pi.\\
\end{align*}

The $\mathrm{P_{high}}$ point serves as the counterpart 
to the $\mathrm{P_{low}}$ point. In contrast to 
the $\mathrm{\Gamma_{high}}$ and $\mathrm{H_{low}}$ points, 
where the upper and lower conical bands exchange their eigenspaces, 
the eigenspaces of the conical bands at the 
$\mathrm{P_{high}}$ and $\mathrm{P_{low}}$ points are 
related by complex conjugation. 
Consequently, the Berry connections and Berry curvatures 
at the $\mathrm{P_{high}}$ point are also complex conjugates 
of those at the $\mathrm{P_{low}}$ point, 
leading to opposite topological charges.

The eigenvalues of the effective Hamiltonian
$H_{\mathrm{P_{high}},\mathrm{eff}}(\bm{k})$ are given by
\begin{align*}
    E_\mathrm{P_{high}}^{\pm} &= \pm \frac{\gamma a}{2\sqrt{3}}k,\\   
\end{align*}
which correspond to a pair of conical bands forming a simple Dirac cone 
at the $\mathrm{P_{high}}$ point, as shown in 
Fig.~\ref{fig:k4_Weyl_points} ($\mathrm{P_{high}}$).

The normalized eigenfunctions are
\begin{align*}
    \bm{u}_\mathrm{P_{high}}^{-} &= 
    \frac{1}{\eta_-}
    \begin{pmatrix}
    \frac{1}{\sqrt{2}}(k_x+k_y+k_z-\sqrt{3}k) \\
    \omega k_x+\omega^2 k_y +k_z
    \end{pmatrix},
    \\
    \bm{u}_\mathrm{P_{high}}^{+} &= 
    \frac{1}{\eta_+}
    \begin{pmatrix}
    \frac{1}{\sqrt{2}}(k_x+k_y+k_z+\sqrt{3}k) \\
    \omega k_x+\omega^2 k_y +k_z
    \end{pmatrix}.\\
\end{align*}

The corresponding Berry connections are
\begin{align*}
    \bm{A}_\mathrm{P_{high}}^{-} &= 
    \frac{1}{2k(k_x+k_y+k_z-\sqrt{3}k)}
    \begin{pmatrix}
    -k_y+k_z\\-k_z+k_x\\-k_x+k_y
    \end{pmatrix},
    \\
    \bm{A}_\mathrm{P_{high}}^{+} &= 
    \frac{1}{2k(k_x+k_y+k_z+\sqrt{3}k)}
    \begin{pmatrix}
    k_y-k_z\\k_z-k_x\\k_x-k_y
    \end{pmatrix},\\
\end{align*}
and the associated Berry curvatures are
\begin{align*}
    \bm{\Omega}_\mathrm{P_{high}}^{-} 
    &=\frac{+\bm{k}}{2k^3},
    &
    \bm{\Omega}_\mathrm{P_{high}}^{+}
    &=\frac{-\bm{k}}{2k^3}.\\
\end{align*}

Accordingly, the topological charges at 
the $\mathrm{P_{high}}$ point are given by
\begin{align*}
    \phi_\mathrm{P_{high}}^{-} &= +2\pi,&
    \phi_\mathrm{P_{high}}^{+} &= -2\pi,\\
\end{align*}
which are the reverse of those found at the $\mathrm{P_{low}}$ point.

\bibliography{reference} 
\end{document}